\documentclass[preprint,5p,number,sort&compress]{elsarticle}
\usepackage{dcolumn}
\usepackage{bm}
\usepackage{upgreek}
\usepackage{graphicx}
\usepackage{hyperref}
\usepackage[load-configurations=abbreviations,detect-all=true]{siunitx}
\usepackage[version=3]{mhchem}
\usepackage[usenames,dvipsnames,svgnames,table]{xcolor} 
\usepackage[normalem]{ulem} 
\usepackage{transparent}

\usepackage{xspace}

\newcommand{\TM}{\emph{Topmetal}\xspace}
\newcommand{\TMI}{\mbox{\emph{Topmetal-I}}\xspace}

\newcommand{\AmAlpha}{\ensuremath{^{241}\text{Am}}\xspace}
\newcommand{\VR}{\ensuremath{V_\text{reset}}\xspace}
\newcommand{\ND}{``Node''\xspace}
\newcommand{\VN}{\ensuremath{V_\text{node}}\xspace}
\newcommand{\Vc}{\ensuremath{\Delta U}\xspace}

\DeclareSIUnit\keV{keV}
\DeclareSIQualifier\ee{ee}
\DeclareSIQualifier\nr{nr}
\DeclareSIUnit\pe{p.e.}

\journal{Nuclear Instruments and Methods in Physics Research A}

\begin{document}

\begin{frontmatter}

\title{Development of a highly pixelated direct charge sensor, \TMI, for ionizing radiation
  imaging}

\author[ccnu]{Yan Fan}
\author[ccnu]{Chaosong Gao}
\author[ccnu]{Guangming Huang}
\author[ccnu]{Xiaoting Li}
\author[lbnl]{Yuan Mei\corref{cor0}}\ead{ymei@lbl.gov}
\author[ccnu]{Hua Pei}
\author[ioa]{Quan Sun}
\author[ccnu]{Xiangming Sun\corref{cor0}}\ead{xmsun@phy.ccnu.edu.cn}
\author[ccnu]{Dong Wang}
\author[ccnu]{Zhen Wang}
\author[ccnu]{Le Xiao}
\author[ccnu]{Ping Yang}

\address[ccnu]{Central China Normal University, Wuhan, Hubei 430079, China}
\address[lbnl]{Nuclear Science Division, Lawrence Berkeley National Laboratory, Berkeley,
  California 94720, USA}
\address[ioa]{Institute of Acoustics, Chinese Academy of Sciences, Beijing 100190, China}

\cortext[cor0]{Corresponding author}

\begin{abstract}
  Using industrial standard \SI{0.35}{\micro\meter} CMOS Integrated Circuit process, we realized
  a highly pixelated sensor that directly collects charge via metal nodes placed on the top of
  each pixel and forms two dimensional images of charge cloud distribution.  The first version,
  \TMI, features a $64\times64$ pixel array of \SI{80}{\micro\meter} pitch size.  Direct charge
  calibration reveals an average capacitance of \SI{210}{fF} per pixel.  The charge collection
  noise is near the thermal noise limit.  With the readout, individual pixel channels exhibit a
  most probable equivalent noise charge of \SI{330}{e^-}.

\end{abstract}

\end{frontmatter}

\section{Introduction}\label{sec:introduction}

Charge sensors are at the heart of many ionizing radiation detectors.  Modern imaging systems
often require a simultaneous determination of the amount, position and time structure of charges
from ionization.  In these cases, a charge sensor of sub-millimeter spatial resolution is highly
desirable.

Solid state detector systems, mainly segmented silicon and germanium crystals coupled to CMOS
pixel sensors, are widely deployed.  CMOS pixel sensors such as
Medipix/Timepix\cite{Ballabriga2011S15,Llopart2007485} and FE-I3/FE-I4 \cite{FE-I4} are readily
available.  They have pixel sizes of tens to hundreds of microns.  They are coupled to the
crystals through processes such as the flip-chip bump bonding.

On the other hand, for gaseous and liquid detectors, wire and Printed Circuit Board (PCB) are
still the most popular readout schemes.  Those detectors are often configured as Time Projection
Chambers (TPCs \cite{TPC}).  Noticeable imaging systems of such kind are LXeGRIT\cite{Aprile1996}
and $\mu$-PIC\cite{Nagayoshi200420}.  For practical reasons, it is difficult to realize a
multi-wire readout with a wire pitch smaller than a millimeter.  Novel designs using PCB have
reached a pitch of hundreds of \si{\micro m} \cite{Nagayoshi200420}.  However, such designs still
face challenges of signal readout multiplexing and speed.

Gaseous and liquid detectors have distinct advantages over solid state detectors: they are more
easily scalable to large mass and are resilient to radiation damage.  Only with an equally
scalable yet high resolution charge readout scheme, their superior properties can be exploited in
imaging applications.  The CMOS pixel sensor is an excellent candidate for this task, because of
the fine pixel size, as well as the possibility of embedding complex circuitry for signal paths.

There have been attempts towards the use of CMOS pixel sensors in a TPC to read the charge
signal, with limited success.  The D$^3$ experiment \cite{d3} uses FE-I3/FE-I4 sensors developed
for the ATLAS \cite{atlas} experiment, behind a gaseous electron multiplication stage, to detect
charge tracks resulting from potential dark matter interactions.  A similar attempt was made to
use the Timepix sensor as the readout for a TPC \cite{GEM-TPC-PXL}.

The development of CMOS sensors is usually dictated by the requirements of large experiments or
the consensus of consortiums.  Existing sensors either have some high level processing already
built-in to handle the massive data rate common in collider experiments, or are lack of low noise
analog channels.  The readout for gaseous and liquid detectors are often slower, but have more
stringent requirement on noise performance.

We set out to realize a CMOS sensor that is uniquely suitable for charge readout in gaseous and
liquid detectors.  We implemented a direct charge sensor with \SI{80}{\micro m} pitch between
pixels using the industrial standard \SI{0.35}{\micro m} CMOS process.  Its behavior in charging
collection is well characterized.

\section{Sensor Structure and Operation}\label{sec:sensor-structure-operation}

\begin{figure}[htbp]
  \centering
  \includegraphics[width=0.5\linewidth]{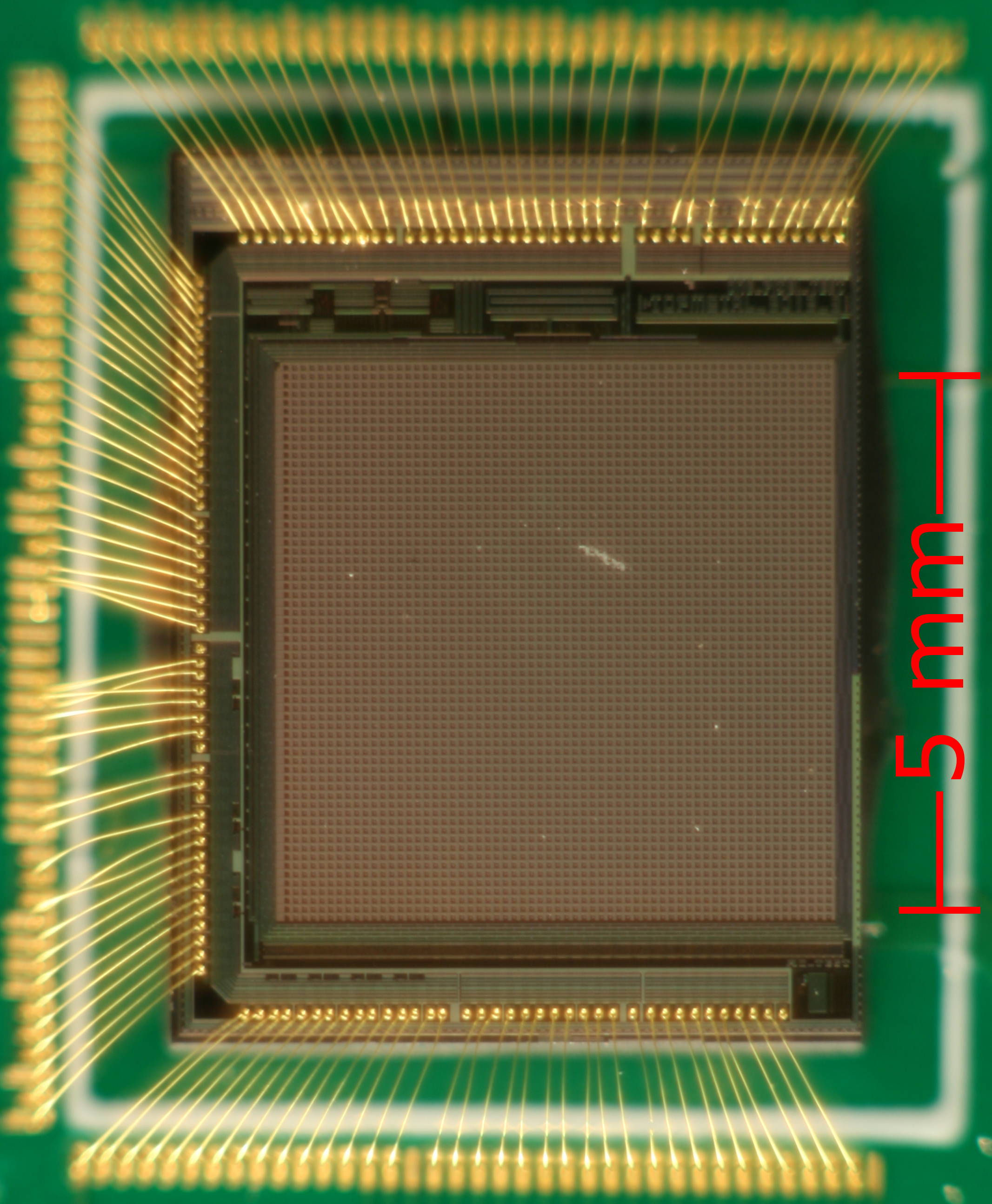}\hspace{2em}%
  \includegraphics[width=0.33\linewidth]{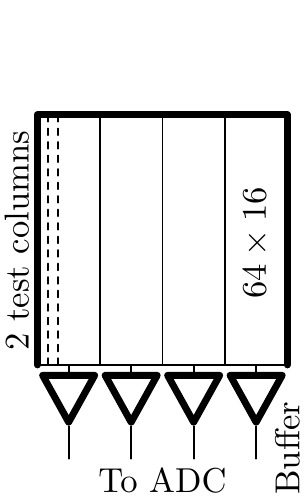}
  \caption{Photograph of a \TMI sensor (left).  The sensor chip is placed on a PCB and gold
    wire-bonded.  The $64\times64$ pixel array, with \SI{80}{\micro m} pitch distance between
    pixels, constitutes an approximately $5\times\SI{5}{mm^2}$ charge sensitive area in the
    center of the sensor.  The sensor is divided into four $64\times16$ sub-arrays with dedicated
    analog buffer inside each sub-array.  Two test pixel columns are implemented in the left-most
    sub-array (right).}
  \label{fig:TMIPhoto}\label{fig:TMIOverallSketch}
\end{figure}

A photograph of one fully fabricated and wire-bonded \TMI sensor is shown in
Fig.~\ref{fig:TMIPhoto}.  The sensor is implemented in a $6\times\SI{8}{mm^2}$ silicon
real-estate area.  With \SI{80}{\micro m} pitch distance between pixels, the $64\times64$ square
pixel array makes up a $5.12\times\SI{5.12}{mm^2}$ charge sensitive region.  Readout interface
logic and analog buffers are placed around the pixel array.

The pixel array is divided into $4$ sub-arrays for efficient analog readout.  As shown in
Fig.~\ref{fig:TMIOverallSketch}, each sub-array is of $64$ rows by $16$ columns in size and has
one dedicated analog output buffer per sub-array.  A single external clock drives the analog
row/column multiplexing readout circuitry in each of the $4$ sub-arrays synchronously.

\begin{figure}[htbp]
  \centering
  \includegraphics[width=0.8\linewidth]{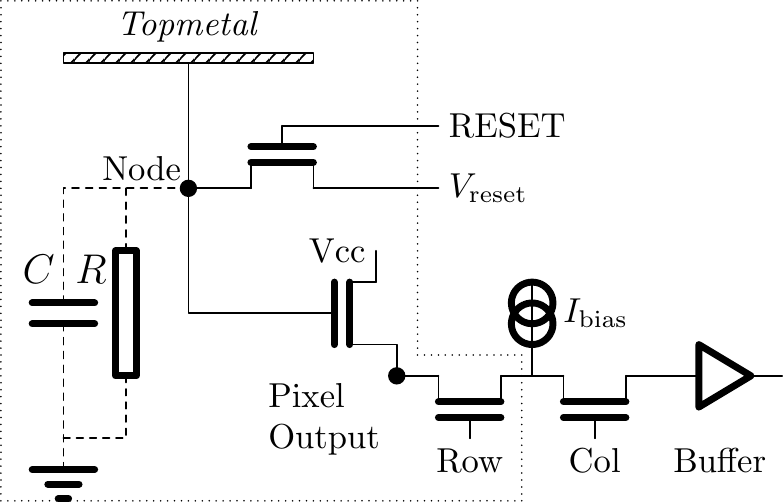}
  \caption{Internal structure of a single pixel.  The \TM is a metal patch exposed to the
    environment external to the sensor to collect charge directly.  The components enclosed in
    the dotted box is unique to each individual pixel, the column selection transistor is shared
    by all the pixels in the same column, and the output buffer is shared by all the pixels in
    the same sub-array.  The capacitor $C$ and the resistor $R$, connected by dashed lines, are
    not real components implemented in the sensor.  They illustrate the equivalent circuitry of
    the \TM.  Its behavior is discussed in the text.  Only the analog channel is shown.}
  \label{fig:TMIPixelSketch}
\end{figure}

The internal structure of a single pixel is illustrated in Fig.~\ref{fig:TMIPixelSketch}.  The
\TM is implemented with the \texttt{PROBEPAD} component in the standard CMOS process, which is a
patch of metal in the topmost layer and has a part of it not covered by the passivation layer.
We designed the metal patch to be $25\times\SI{25}{\micro m^2}$ in size, with the central
$15\times\SI{15}{\micro m^2}$ exposed hence sensitive to external charge.  The metal patch is
large enough to be mechanically and chemically stable, while small enough to suppress the
cross-talk between neighboring pixels to a negligible level.

Charges arrived at the sensor and collected by the \TM result in an electric potential change at
the \ND, which is a detectable signal.  The potential of the \ND, \VN, is captured by means of a
source follower that drives the signal through row/column multiplexer for analog readout.  The
equivalent capacitance of \TM to ground is measured to be approximately \SI{210}{fF}.  It means
$1000$ electrons collected by \TM would correspond to \SI{0.8}{mV} voltage drop on the \ND.

The resetting circuit for the \ND allows controlled removal of charge collected by the \TM and
the initial potential setting of the node.  When a logic high is applied to the reset gate, the
\ND potential is set to \VR and any excess charge on the \ND is neutralized by the power supply
driving \VR.

Two test columns are implemented in the leftmost sub-array (Fig.~\ref{fig:TMIOverallSketch}).
Pixels in the first test column have identical structure as the rest of the pixels, except that
the \ND is internally connected to ground.  They serve as markers in the multiplexed readout.
Since their \VN is fixed internally, they are immune to the noise generated by the \TM
capacitance.  Noise measurements on the first test column measures the noise due to the analog
readout path only, decoupled from that from the \TM.  Pixels in the second test column have the
\ND capacitively coupled to an external pin.  They were designed to test the pixel response to
pulses injected from the external pin.  The pin is tied to ground during normal operations.

Charge signal in each pixel is also converted into digital values by means of a threshold
discriminator (comparator), a delayed resetter, and a 5-bit counter.  Muxing readout for digital
data at array level is also implemented.  The digital data chain is suitable for charge event
counting applications.  However, in this letter, we focus on the physical behavior and analog
characteristics of the device.

\section{Signal Formation}\label{sec:signal-formation}

The \TM can be modeled as a resistor with large resistance $R$ and a capacitor with small
capacitance $C$ connected in parallel (Fig.~\ref{fig:TMIPixelSketch}).  The time constant
$\tau=RC$ is on the order of \SI{10}{seconds}.  During charge collection, the equivalent charge
current $i$ will run through the resistor while charging up the capacitor at the same time.  The
following differential equation gives the general behavior of the circuitry:
\begin{equation}
  \label{eq:1}
  i(t)=\frac{U}{R}+C\frac{dU}{dt}\;.  
\end{equation}
$U$ is the voltage across the capacitor, which is then measured.

The solution to the differential equation is 
\begin{multline}
  \label{eq:2}
  U(t)=\\
  \begin{cases}
    U_0e^{-t/\tau} & i=0\\
    (U_0-iR)e^{-t/\tau}+iR & i=\text{constant}\\
    \displaystyle e^{-t/\tau}\left(U_0+\frac{1}{C}\int_0^t i(x)e^{x/\tau}dx\right) & i(t)
  \end{cases}
\end{multline}
The time-dependent $i(t)$ solution is the most generic.  The solutions for $i=0$ and
$i=\text{constant}$ are two special cases.  The boundary conditions are set such that no matter
whether there is a charge current $i$ or not, at $t=0$, the output voltage $U(t=0)$ is set to a
defined reset value $U_0$.  When collecting electrons (negative charge), $i<0$, otherwise $i>0$.

\begin{figure}[htbp]
  \centering
  \includegraphics[width=0.9\linewidth]{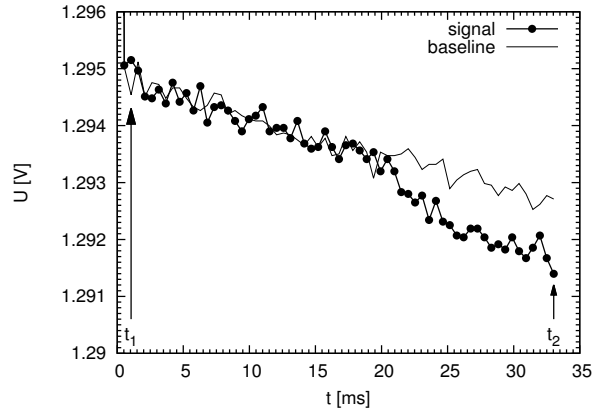}
  \caption{Temporal behavior of the \ND voltage in a pixel.  One example in the real data.  Reset
    happens at $t=0$.  The node voltage at reset is about \SI{1.4}{V} (off scale).  The curve
    with filled circles shows node voltage during charge collection.  A cluster of charge
    suddenly arrives at $t_q\approx\SI{21}{ms}$.  The thin curve is the baseline, showing the
    voltage is decaying exponentially with a large time constant.}
  \label{fig:vdecay}
\end{figure}

The sudden arrival of a charge pulse can be modeled as $i(t)=Q\delta(t-t_q)$, where $t_q$ is the
arrival time of the charge pulse.  At $t=t_q$, the voltage decay curve $U(t)$ drops by the amount
of $Q/C$, as shown in Fig.~\ref{fig:vdecay}.

Ideally, we can build an algorithm to find the sudden drops and reconstruct their amplitudes to
measure the charge collected.  However, due to the noise, as seen in Fig.~\ref{fig:vdecay}, such
an approach is unlikely to be robust.  Also, in the case of continuous charge current, there will
be no sudden drop.  Instead, to extract the charge information, we employ a ``double
subtraction'' scheme.

Prior to the charge measurement, we remove the charge source, but operate the sensor in the same
condition, to establish a baseline.  The baseline, $U(t)|_{i=0}$, is a time-series measurement of
node voltage without charge input.  During the charge measurement, we record $U(t)|_i$ in a
similar fashion.  The baseline and charge measurements are synchronized by a periodical reset
signal.  The time between two resets defines an event window, and $t_2-t_1$ is the charge
integration time.  We subtract the voltage right after reset, $U(t_1)$, from the voltage at the
end of the window, $U(t_2)$, in each case, then subtract the difference in baseline from that in
charge measurement.  The quantity \Vc is shown in Eq.~\eqref{eq:3}.
\begin{equation}
  \label{eq:3}
  \begin{aligned}
    \Delta U&=\left[U(t_2)-U(t_1)\right]_{i}-\left[U(t_2)-U(t_1)\right]_{i=0}\\
            &=-iR\left(e^{-t_2/\tau}-e^{-t_1/\tau}\right)\,.
  \end{aligned}
\end{equation}
In the limit where $t/\tau\rightarrow0$,
\begin{equation}
  \label{eq:4}
  e^{-t_2/\tau}-e^{-t_1/\tau}\approx-(t_2-t_1)/\tau\,,
\end{equation}
hence
\begin{equation}
  \label{eq:5}
\Delta U\approx\frac{i(t_2-t_1)R}{RC}=\frac{Q}{C}\,.
\end{equation}
Eqs.~(\ref{eq:3}--\ref{eq:5}) are derived assuming a constant current $i$.  It can be shown that
the same relation Eq.~\eqref{eq:5} holds for sudden charge arrivals.

We operate at a charge integration time window of tens of milliseconds, so that Eq.~\eqref{eq:4}
is satisfied.  Therefore we can use the simple proportional relation Eq.~\eqref{eq:5} to compute
the charge value.  We also refer to one \Vc measurement an ``event''.

The node voltage behavior in a characteristic pixel is shown in Fig.~\ref{fig:vdecay}.  When the
reset is activated ($t=0$), the node is set to the voltage of \VR.  As soon as the reset is
deactivated, due to charge injection on the MOS transistor \cite{charge_injection}, the node
voltage drops to a deterministic value.  In this example, the node voltage right after reset is
about \SI{1.295}{V}.  Up to a small fluctuation, the baseline curve behaves identically after
each reset for the same pixel.  Then, when there is no charge being collected, the node voltage
changes exponentially towards a stable value.  When the charges arrive, the voltage curve shows a
sudden drop.

Due to the inhomogeneity of the CMOS process, the baseline curves behave differently among
pixels.  The most significant variation is in the decay time constant.  Nevertheless, it is only
necessary to calibrate the baseline curves for every pixel once prior to the charge detection.

To image 2D distributions of the charge cloud, the above procedure is performed on each pixel in
the array.  The temporal behavior of each pixel is decoded from the multiplexed readout signal.
The interval of sampling is the pixel clock period times the number of pixels in a sub-array
(1024).  The time between resets is set to be slightly larger than the desired charge integration
time.  \Vc is then computed for each pixel and the 2D distribution is established.

\section{Charge Calibration}\label{sec:charge-calibration}

To demonstrate that the \TMI sensor is capable of directly detecting charge, and to calibrate for
charge measurements, we constructed a rudimentary drift chamber, using an \AmAlpha alpha source
to ionize air, then drift the negative charge towards the sensor in an electric field.

\begin{figure}[htbp]
  \centering
  \includegraphics[width=0.5\linewidth]{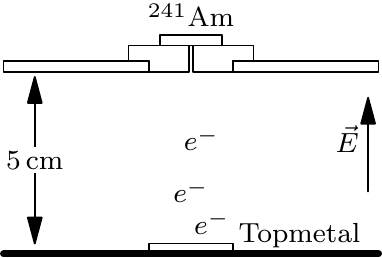}
  \caption{Setup for \TMI sensor charge calibration.  Not drawn to scale.  The sensor is placed
    on a PCB.  They are on an electrical potential within \SI{3.3}{V} of ground.  An \AmAlpha
    source, its collimator and a metal plate are held at negative high voltage and placed above
    the \TMI sensor, parallel to the PCB.  An electric field of \SI{200}{V/cm} is created between
    the metal plate and the PCB.  Alphas from the \AmAlpha source ionize air and the negative
    charge is drifted towards the sensor.}
  \label{fig:TMITestSetup}
\end{figure}

We wire-bonded a \TMI sensor onto a PCB and placed supporting circuity on the PCB to drive the
sensor and to transmit analog signals out via coax cables.  On the other end of the coax cables,
signals are digitized by a 14-bit high speed ADC.  To create a uniform drift field with field
lines perpendicular to the sensor, we placed a large aluminum plate \SI{5}{cm} above, and
parallel to the sensor, serving as the cathode.  While the sensor substrate and the PCB are
placed close to ground potential, the aluminum plate is held at \SI{-1000}{V}, creating an
electric field of \SI{200}{V/cm}.

\begin{figure}[htbp]
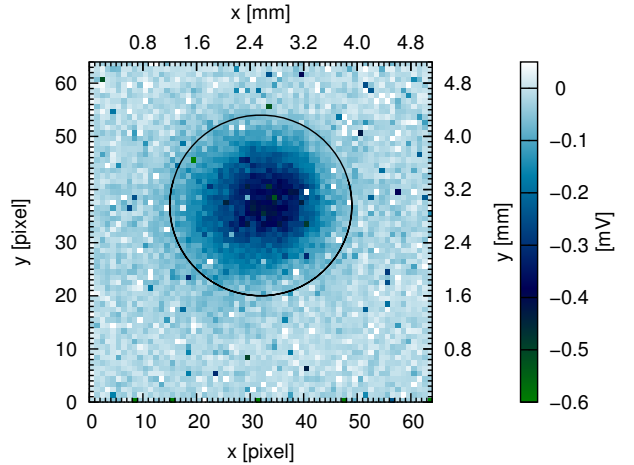

  \centering
  \includegraphics[width=0.99\linewidth]{{{fig/img_sums1_lt_-0.1}}}
  \caption{Charge cloud distribution imaged by the \TMI sensor.  The image is an average of
    events mostly containing one alpha in each event.  Each pixel is \SI{80}{\micro m} in size.
    Alpha particles traverse perpendicular to the sensor surface, but are stopped in air before
    hitting the sensor.  The charge is then drifted towards the sensor and collected.  The
    integration time is \SI{32}{ms} for each event.}
  \label{fig:ArrayAlphaImg}
\end{figure}

A collimated \AmAlpha source embedded in the cathode emits alpha particles perpendicular to the
sensor.  The alphas ionize air along their paths.  Negative charge generated during the process
is drifted towards the sensor and eventually collected by the sensor.  Alphas from the \AmAlpha
source have a range of about \SI{4}{cm} in air at standard temperature and pressure \cite{astar}.
The distance between the cathode and the sensor (\SI{5}{cm}) was chosen so that alphas would
deposit their full energy in air instead of hitting the sensor.  An averaged image of the charge
cloud from alphas seen by the sensor is shown in Fig.~\ref{fig:ArrayAlphaImg}.

The event time window is \SI{33.5}{ms} and the integration time is \SI{32}{ms}.  Effectively we
are collecting data at an event rate of \SI{30}{Hz}.  The collimator limits the alpha rate to be
much lower than the data event rate.  The conditions are chosen so that within a data event,
there can be either zero or one alpha, and the probability of two or more alphas in the same
event becomes negligible.  Meanwhile, the integration time is long enough to fully collect the
charge after an alpha interaction, since it takes only microseconds for electrons to drift for
\SI{5}{cm} \cite{edrift_vel_air}.

\begin{figure}[htbp]
  \centering
  \includegraphics[width=0.9\linewidth]{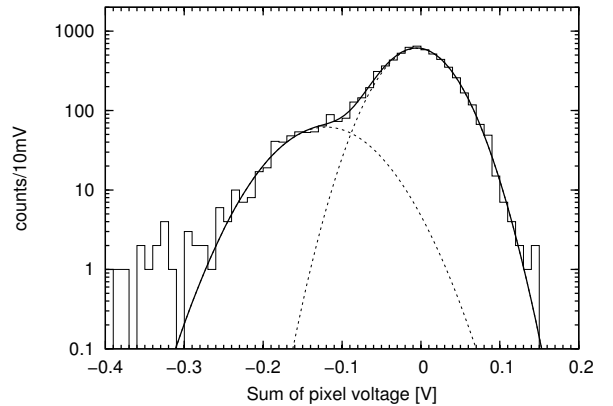}
  \caption{Voltage summation of pixels within the selection circle.  A sum of two Gaussian
    functions (thick curve) is used to fit the histogram.  The dashed lines show the constituents
    of the fit function.  The fit yields a $\chi^2/\text{NDF}=53.8/60$.  Events containing no
    alpha fill the high peak centered around 0.  The mean of the single-alpha peak is
    \SI[separate-uncertainty]{-0.120+-0.008}{V}}
  \label{fig:sums1hist}
\end{figure}

On average, each alpha particle emitted from the \AmAlpha source deposits \SI{5.45}{MeV}
\cite{am241alphaspec} energy in air resulting in ionization.  The air has an ionization $W$-value
of \SI{35}{eV} \cite{w_val_air}.  Since alphas are fully stopped before hitting the sensor, and
electrons are fully collected during the integration time, each alpha event results in a mean
value of approximately \SI{1.56e5}{e^-}.  Since each event has either zero or one alpha, if we
plot a histogram of the voltage signal due to charge, we should be able to identify a peak
corresponding to single alphas, and use the relation
\begin{equation}
  \label{eq:6}
  C = Q\Big/\sum_{mn}\Delta U_{mn}  
\end{equation}
to compute the capacitance $C$ of a single pixel.  $Q$ is the total charge and the summation is
over pixels that see charge.

\Vc's of pixels within a \SI{17}{pixel} radius (Fig.~\ref{fig:ArrayAlphaImg}) are summed up to
fill the histogram in Fig.~\ref{fig:sums1hist}.  The region is chosen such that \SI{90}{\percent}
of charge is contained in the region given the signal is modeled by a two-dimensional Gaussian
distribution.  We will not attempt to correct for the potential loss of \SI{10}{\percent} charge,
instead, we treat it as a systematic uncertainty of the measurement.  The single-alpha peak is
clear in Fig.~\ref{fig:sums1hist} and is well characterized by a Gaussian function.  From the
mean voltage value of the single-alpha peak, we measure the pixel capacitance $C=\SI{207}{fF}$.
The uncertainty, a combination of uncertainty from the fit and the systematic uncertainty, is
found to be \SI{12}{\percent}.

To validate the result, we used a less stringent collimator to increase the total ionization per
unit time.  Under the same electric field, we measured the ionization current with a picoammeter
to be around \SI{10}{pA}.  The ionization current is then collected by the \TMI sensor operating
under exactly the same condition.  The measurement yields a similar capacitance value, with
larger uncertainty.

\begin{figure}[htbp]
  \centering
  \includegraphics[width=0.9\linewidth]{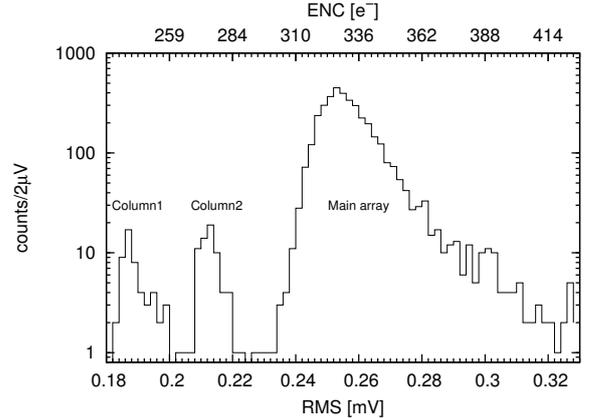}
  \caption{RMS fluctuation distribution of pixels.  The left most two peaks correspond to test
    pixels in the first and second column respectively.  ENC is computed assuming
    $C=\SI{207}{fF}$.}
  \label{fig:pixelBlRMS}
\end{figure}

To evaluate the noise performance of the sensor, we use the baseline dataset alone, and compute
the Root-Mean-Square (RMS) of voltage drop $\left[U(t_2)-U(t_1)\right]_{i=0}$ of each pixel
across all events.  The noise is then calculated as
$C\times\text{RMS}(\left[U(t_2)-U(t_1)\right]_{i=0})$, where $C$ is the above measured pixel
capacitance.  The RMS voltage drop distribution of the entire pixel array is shown in
Fig.~\ref{fig:pixelBlRMS}.  The distribution has three distinct clusters.  The cluster with the
lowest RMS value contains the test pixels in the first column.  Those pixels have their voltages
tied to a fixed value at all times.  The RMS value of them comes solely from the analog readout
noise.  It provides a measurement of the analog readout noise independent of the noise from the
\TM capacitance.  The second test column is capacitively coupled to an external pin therefore has
higher noise.  The RMS of the main array has its most probable value at \SI{0.253}{mV}, which
corresponds to about \SI{327}{e^-}.  The uncertainty of this value is solely determined by the
uncertainty of capacitance measurement.

During normal air-cooled operations, the sensor temperature stabilizes at about
\SI{57}{\degreeCelsius}.  A \SI{207}{fF} capacitor at such temperature has an intrinsic thermal
noise of \SI{192}{e^-} \cite{kTC_noise}.  Using the first test pixel column, we determine that
the analog readout system has an ENC of \SI{254}{e^-}.  The quadratic summation of the two noise
values yields \SI{318}{e^-}, which is very close to the most probable ENC value of the main pixel
array.  It indicates that the equivalent capacitor of the \TM is operating near its thermal noise
limit.

\section{Summary}\label{sec:summary}

We have demonstrated the possibility of implementing a highly pixelated sensor for direct charge
collection using industrial standard \SI{0.35}{\micro m} CMOS technology.  No post-processing is
necessary.  Although in this first attempt, \TMI, the sensor noise is moderately high, its
capability in observing charge from single-alpha events in air directly already shows its
potential in applications.

The design and production of CMOS IC are traditionally perceived as inhibitively expensive.
However, recent advancement in the industry has made the mature technologies such as the
\SI{0.35}{\micro m} process affordable.  Also, the direct charge collection capability, without
the need of any post-processing, makes the overall integration cost much lower than for example
technologies used in producing hybrid solid-state detectors.

To improve beyond \TMI, besides reducing the pixel size, we can further reduce the noise and
record the charge arrival time.  By placing a charge sensitive amplifier in each pixel, and
reducing the \TM size hence its capacitance, we foresee an ENC better than \SI{10}{e^-} per pixel
should be achievable.  Information on charge arrival time can be obtained using an individually
addressable pixel structure.  We will explore these options in future series of \TM sensors.

\section*{Acknowledgments}

This work is supported, in part, by the Thousand Talents Program at Central China Normal
University and by the National Natural Science Foundation of China under Grant No.~11375073.  We
also acknowledge the support from LBNL for hosting the physical measurements of the sensor.  We
would like to thank Christine Hu-Guo and Nu Xu for fruitful discussions.



\bibliographystyle{elsarticle-num-names}
\bibliography{refs}

\end{document}